\documentclass[%
 reprint,
superscriptaddress,
 amsmath,amssymb,
 aps,
]{revtex4-1}

\usepackage{graphicx}% Include figure files
\usepackage{dcolumn}% Align table columns on decimal point
\usepackage{bm}% bold math
\usepackage[colorlinks,allcolors=blue]{hyperref}

\begin{document}

\preprint{APS/123-QED}

\title{Attosecond Polarisation Modulation of X-ray Radiation in a Free Electron Laser}% Force line breaks with \\

\author{J. Morgan}
 \email{jenny.morgan@strath.ac.uk}
 \affiliation{%
University of Strathclyde (SUPA), Glasgow G4 0NG, United Kingdom\\
}
 \affiliation{%
Cockcroft Institute, Warrington, WA4 4AD, UK\\
}

\author{B. W. J. M$^{\rm c}$Neil}
 \affiliation{%
University of Strathclyde (SUPA), Glasgow G4 0NG, United Kingdom\\
}
 \affiliation{%
Cockcroft Institute, Warrington, WA4 4AD, UK\\
}
\affiliation{ASTeC, STFC Daresbury Laboratory, Warrington, WA4 4AD, UK}

\begin{abstract}
    A new method to generate short wavelength Free Electron Laser output with modulated polarisation at attosecond timescales is presented. Simulations demonstrate polarisation switching timescales that are four orders of magnitude faster than the current state of the art and, at X-Ray wavelengths, approaching the atomic unit of time of approximately $24$~attoseconds. Such polarisation control has significant potential in the study of ultra-fast atomic and molecular processes. The  output alternates between either orthogonal linear or circularly polarised light without the need for any polarising optical elements. This facilitates operation at the high brightness X-ray wavelengths associated with FELs. As the method uses an afterburner configuration it would be relatively easy to install at exciting FEL facilities, greatly expanding their research capability.

\end{abstract}

\maketitle

%\section{Introduction}

The polarisation of light is a fundamental property which  affects its interactions with matter. These interactions are used experimentally to investigate various properties of matter. Current experiments can demand a greater control and flexibility of the polarisation than the generation of purely circular, elliptical or linearly polarised light. In particular, temporal switching of light polarisation is desirable. If this can be done  at the fast timescales comparable to those of atomic processes, it can enable experimental investigation of these processes. For example, control over the handedness of circular polarisation may be useful in the study of spin related processes\cite{PhysRevB.93.155426, nicholls2017ultrafast}. 
However, such fast switching of the polarisation properties of light is a non-trivial task as conventional polarising elements are quasi-static devices at these timescales. 
While some conventional polarising elements can be controlled by electric currents \cite{10.1117/12.567640}, these are limited by their electronic components to gigahertz switching speeds and also see large energy losses.  

%Novel techniques have improved on this switching rate. $800$fs  linear polarisation switching rate is achieved using switchable reflective polarizer based on plasmonic technology. Control over the handedness, and therefore photon spin direction, of circular polarisation is yet more difficult with    

Some recent methods have improved on this switching rate via the use of plasmonic technologies to produce linear polarisation switching within 800fs~\cite{yang2017femtosecond}, and circular polarisation switching at pico-second timescales~\cite{nicholls2017ultrafast}. However, these techniques are based on the active control of polarising elements and operate primarily at visible wavelengths or longer.  As wavelengths shorten below that of the ultraviolet, polarising optics are more limited and the radiation polarisation is primarily determined by the method of generation. 

In electron accelerator based light sources, which can generate light into the hard X-ray, the motion of the radiating electrons propagating though magnetic undulators determines the polarisation of the photon beam. For example, circular polarisation modulation has been demonstrated in a  synchrotron by controlling electron bunch orbits through twin undulators~\cite{holldack2020flipping}. Although this modulates the X-ray polarisation, the modulation rate is limited to less than $500$MHz. 

In this Letter a method is described which could improve the polarisation switching rate of both linear and circularly polarised high brighness X-rays towards the attosecond timescale regime. This is comparable to the period of for a ground state electron in the Bohr hydrogen atom, the atomic unit of time $\approx 24$as. 

The method modulates the polarisation of light radiated from a Free Electron Laser.  
Trains of radiation pulses are generated in which each pulse alternates between orthogonal linear or circular polarisation states. FELs are widely tuneable devices operating down to the hard X-ray wavelength range~\cite{nphoton} and the method reported here should be applicable over this full range.
Simulations are carried out here in the soft X-ray generating radiation pulse trains with alternate orthogonal  polarisation pulse timescales of  tens of attoseconds. This is approximately four orders of magnitude faster than any other method \cite{yang2017femtosecond} and eight orders of magnitude faster than polarisation switching available at X-ray wavelengths \cite{holldack2020flipping}. 

\begin{figure*}
\centering
\includegraphics[width=1.0\textwidth]{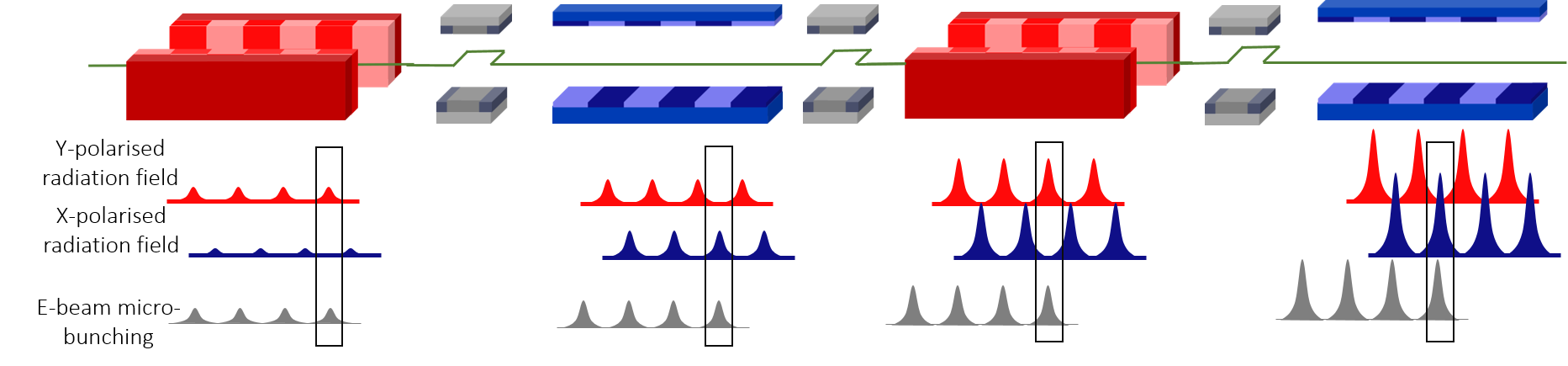}
\caption{\label{fig:schematic} Schematic layout of a section of afterburner used to generate a radiation pulse train with alternating  $x$ and $y$ linear polarisation. In each undulator, those regions  of the electron beam with modulated micro-bunching emit coherently. Chicanes delay the electron beam between undulator modules so that those sections of high micro-bunching overlap with the appropriately polarised pulse for the undulator in which they are propagating. 
}
\end{figure*}

In a high-gain FEL a relativistic electron beam propagates through magnetic undulators and emits electromagnetic radiation (light) with a resonant wavelength $\lambda_r=\lambda_u (1+\overline{a}_u^2)/2\gamma_0^2$, where $\lambda_u$ is the undulator period, $\overline{a}_u$ is the rms undulator parameter and $\gamma_0$ is the electron beam's relativistic factor. The light is amplified via a collective interaction which causes the electrons to micro-bunch at the resonant wavelength and to emit coherently~\cite{nphoton}. The initial non-uniform phase distribution of electrons, or shot-noise, can provide the initial seed which is subsequently amplified in the process of Self Amplified Spontaneous Emission, giving a temporally noisy output~\cite{SASE}. The relative propagation of a radiation wavefront through the electrons of one $\lambda_r$ every $\lambda_u$, referred to as `slippage', allows interaction between different regions of the electron bunch and radiation pulse. This correlates the phase of the radiation output at a length determined by the cooperation length $l_c=\lambda_r/4\pi\rho$ - the relative slippage in an exponential gain length through the undulator $l_g=\lambda_u/4\pi\rho$. Here $\rho$, the FEL parameter, determines the strength of the FEL interaction~\cite{SASE}.  

The FEL process generates high-power radiation with its polarisation determined by the magnetic undulator field - either planar, elliptical or helical. A typical X-ray FEL facility uses planar undulators to microbunch the electrons. Polarisation control can be enabled by adding additional undulators placed downstream of the main planar undulator amplification section once micro-bunching has been established \cite{lutman2016polarization}. Such additional downstream undulators, or `afterburners', are increasingly being explored as a method to tailor FEL output in many ways not limited to polarisation control, e.g. short pulse generation \cite{PhysRevLett.100.203901, dunning2013few} and transverse phase manipulation \cite{hemsing2020coherent}. These afterburners can provide  solutions to meet specific experimental output requirements with minimal changes to an existing facility and therefore at relatively low cost.

To generate FEL output with modulated polarisation, we propose an afterburner design consisting of a series of few period, alternate orthogonaly polarised undulator modules as shown in Fig.~\ref{fig:schematic}. The undulators are separated by electron delay chicanes which can introduce additional slippage between the electron bunch and the radiation field. Both of the orthogonal, polarised radiation fields emitted in the afterburner are mode-locked which creates trains of short pulses~\cite{PhysRevLett.100.203901}. The orthogonally polarised pulse trains are shifted temporally with respect to each other, so that the combined pulse train consists of a series of alternate, orthogonally polarised pulses.

Mode-locking in a FEL, first proposed in~\cite{PhysRevLett.100.203901} and compacted into an mode-locked afterburner configuration in~\cite{dunning2013few}, creates trains of short radiation pulses via a process analogous with mode-locking in conventional cavity lasers~\cite{siegman1986lasers}.
In the mode-locked afterburner, normal FEL amplification occurs first in an electron beam prepared with an energy modulation, $\gamma(t)=\gamma_0+\gamma_m\cos({\omega_m t})$. This generates a periodic micro-bunching structure in the electron beam at the energy modulation period by creating higher FEL gain at the minima of the energy modulated beam.  Chicane delays between the short undulator sections in the afterburner then map the electron beam micro-bunching comb onto the radiation modal pulses.

Here, a similar mapping of the micro-bunched comb to the mode-locked radiation generated in the orthogonally polarised afterburner modules is used to generate alternately polarised pulse trains. Fig.~\ref{fig:schematic} shows a schematic of how these pulse trains are generated in a planar undulator afterburner. Chicanes placed between undulator modules are chosen to delay the high micro-bunched regions of the electron beam to the polarised radiation pulses corresponding to the similarly polarised undulator in which they interact. The orthogonally polarised radiation pulses do not interact with the electrons in this undulator module so that they simply experience free propagation. The orthogonally polarised undulators then effectively behave as additional alternate chicane delays. 

%Chicanes delaying the electron beam control the relative position of electron beam sections with high micro-bunching and the radiation field depending on the type of undulator it propagates through. Radiation does not interact with electrons in undulators with orthogonal polarisation and thus undulators act like chicane delays to the radiation they don't emit. 

The combined slippage of the electrons with respect to a radiation wavefront between undulator modules of the same polarisation should therefore be the modulation period $\lambda_m$. 
%Between undulator modules of the same type, the electrons slip back - due the combined slippage of the chicanes and undulator modules - by the modulation period, $\lambda_m$, so that radiation emitted by a section of bunched electron beam from one undulator overlaps with the succeeding bunched section on entering the next undulator which produces the same polarisation. 
%The slippage between the different undulators is less than $\lambda_m$, and there is no overlap between sections high electron bunching and the orthogonal polarised radiation from the previous undulator.  
The temporal separation of the pulses of radiation with the same polarisation is then $T=\lambda_m/c$ and the relative times of these pulses is:
\begin{equation}
    t_1=nT
\end{equation}
The orthogonally polarised pulses will then have pulse peaks at relative times:
\begin{equation}
    t_2=t_0+t_1+\Delta T
\end{equation}
where $\Delta T=s/c$ is the time for for the radiation to propagate the slippage length, $s= \lambda_m/2$, through the electron bunch between undulator modules of the same polarisation and $t_0$ is a constant which may shift the radiation pulse trains relative to each other. Here, we chose $t_0=0$ so that there is equal spacing between all pulses. 

The method is modeled using the FEL simulation code PUFFIN~\cite{campbell2012puffin} using the parameters based on the LCLS-II project at SLAC~\cite{LCLSIIScience} as listed in Table~\ref{tab:simulationparamers}. Dispersion effects within the chicanes are included in the model although chicanes which reduce dispersion and dispersionless chicanes are being developed~\cite{Clarke:2012zzb, thompson:fel2019-thp033}. 
\begin{table}[htb]
 \centering
\caption{\label{tab:simulationparamers} Simulation Parameters
}
\begin{tabular}{ l @{\qquad} c c}
\toprule
\textrm{Parameter}&
\textrm{Value}\\
\colrule
\textit{Amplifier Stage} \\
Electron beam energy [GeV] & 4\\
Peak current, $I_0$ [kA] & 1\\
rms energy spread $\sigma_{\gamma}/\gamma$ & 0.0125\\
Normalized emittance [mm-mrad] & 0.45\\
rms beam size $\sigma_x$ [$\mu$m] & 26\\
Undulator period $\lambda_{u}$[cm] & 3.9  \\
Resonant wavelength $\lambda_{r}$ [\text{nm}]  & {1.25}\\
Modulation wavelength $\lambda_m$[nm] & {40.0}\\
Modulation amplitude ${\gamma_m}/{\gamma_0}$ & 0.0012\\
 rms  undulator  parameter $\overline{a}_u$  & 1.72\\
$\rho$ parameter & 0.0012\\
\\
\\
\textit{Afterburner}   \\
Number of undulator periods per module & 8\\
Chicane Delays [nm] & 10\\
Number of undulator modules & 36\\
% rms  undulator  parameter $\overline{a}_u$  & 1.72\\

\toprule
\end{tabular}
\end{table}

The method is first demonstrated using an afterburner with alternating $x$ and $y$ planar undulators that will emit correspondingly linearly polarised light. The electron beam is prepared with an energy modulation of period $\lambda_m=40$ nm $=32\lambda_r$ and the electron micro-bunching comb is first developed in a SASE FEL `pre-amplifier' using an $x$-polarised undulator similar to that found at most current FEL facilities. 
The power growth in this pre-amplifier stage is inhibited by the electron beam energy modulation. On subsequent injection into the afterburner, the power growth in the pulsed regions becomes exponential due to their overlap with the high quality electron beam regions being maintained. There is therefore much greater radiation power generated in the afterburner than in the pre-amplifier. The point at which the electron beam is extracted from the pre-amplifier is chosen such that the radiation is two orders of magnitude smaller than the final saturated radiation power in the following afterburner.

Both the $x$ and orthogonal $y$ polarised undulator modules in the afterburner are 8 periods long, each separated by a chicane that delays the electrons by a further 8 resonant wavelengths. The total electron delay is then $s=16\lambda_r=\lambda_m/2$ between successive undulator modules and $\lambda_m$ between undulators of the same polarisation. This maintains overlap between the electron micro-bunching comb and the alternating orthogonaly polarised radiation as shown in Fig.~\ref{fig:schematic}, leading to the amplification of radiation spikes.
%A pulse train structure develops in both the x and y polarised fields. 
The orthogonal radiation spikes so generated should not interfere with each other due to their orthogonal polarisation. However, as both fields are emitted by the same electron beam source, which sees only small changes between undulator modules, fluctuations in the power of one pulse train envelope should be similar to its orthogonal counterpart. 
%A section of micro-bunching radiating a x-polarised undulator will then radiate in the next y-polarised undulator so the pulses will have near identical powers

\begin{figure}[htb]
\centering
\includegraphics[width=0.5\textwidth]{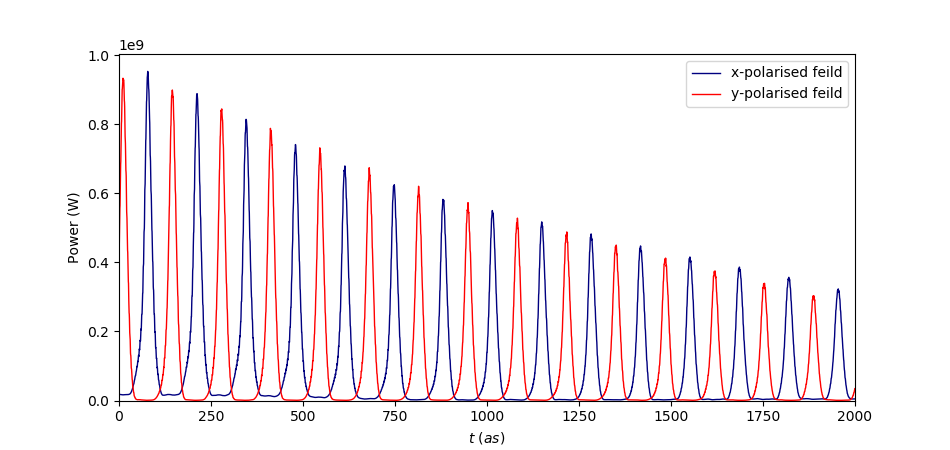}
  \begin{minipage}[b]{0.24\textwidth}
  \centering
    \includegraphics[width=\textwidth]{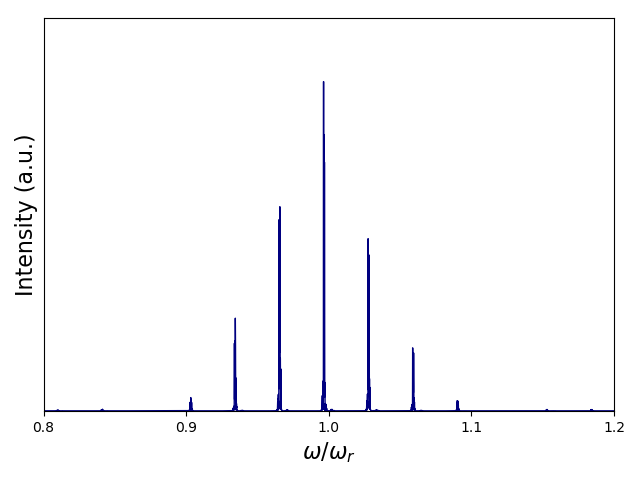}
  \end{minipage}\hfill
  \begin{minipage}[b]{0.24\textwidth}
  \centering
\includegraphics[width=\textwidth]{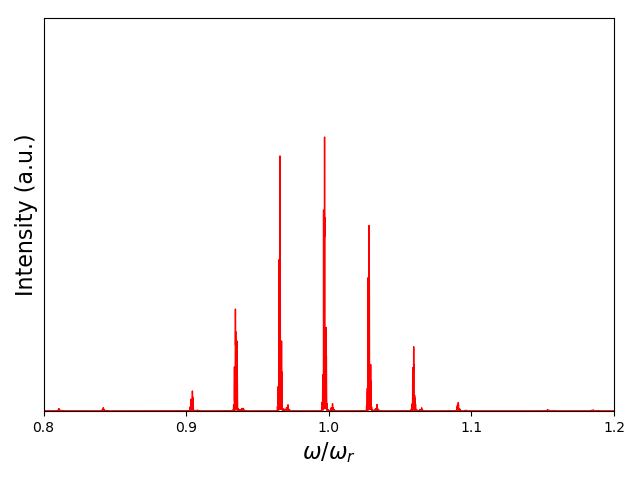}
 \end{minipage}
\caption{\label{fig:energy_feild_overlap} (Top) Power Vs relative time $t$ for the $x$ and $y$ polarised fields and (bottom) the corresponsing spectra after 36 undulator-chicane modules.}
\end{figure}

Fig.~\ref{fig:energy_feild_overlap} shows a section of the radiation power profiles and spectrum of the $x$ and $y$ polarised fields after 36 afterburner undulator modules (16 of each polarisation). The additional slippage between undulator modules leads to a frequency spectrum that is broader than typical FEL output and discretied into frequency modes with modes spacing, $\Delta\omega_s$, as determined by the time taken for the radiation to travel the total slippage length between the same polarised undulators. The radiation pulse peaks arise from the  constructive interference between the frequency modes whose phase has been fixed by the modulation, $\Delta\omega_m=\Delta\omega_s$. This is the principle of mode-locking as described in~\cite{PhysRevLett.100.203901, siegman1986lasers}.

As the undulator modules have equal lengths, both the $x$ and $y$ polarised fields have approximately the same pulse FWHM duration of  $\tau_p\approx 19$~as and with peak powers of $P_{pk}\approx 1$~GW. The separation between each pulse is apprximately $67$ as corresponding to a polarisation switching rate of $15$~PHz.

\begin{figure}[htb]
\centering
\includegraphics[width=0.49\textwidth]{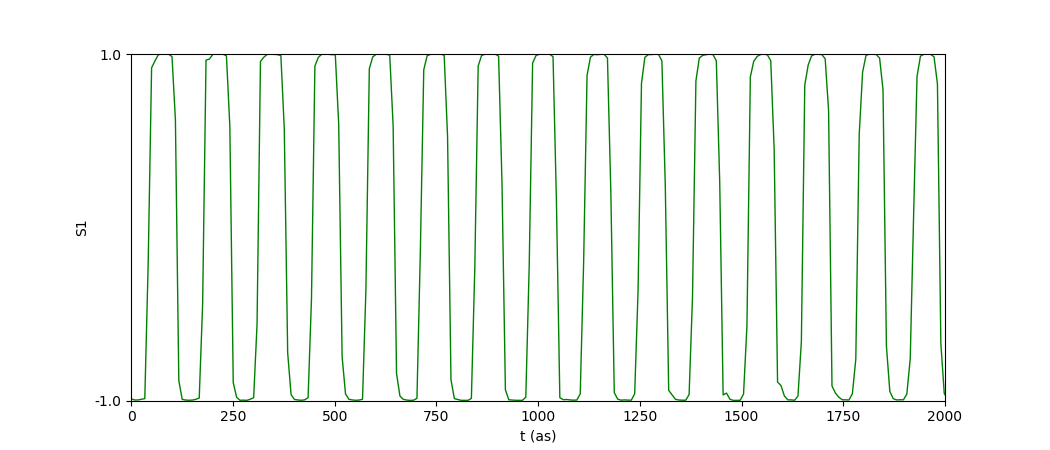}  
\caption{\label{fig:Stokes} The on-axis normalised Stokes parameter $s_1$ as a function of relative time $t$ after 36 undulator-chicane modules. It is seen that $s_1$ flips between positive and negative values with extremes at $|s_1|>0.95$, indicating high degree of polarisation modulation.}
\end{figure}

A normalised Stokes parameter, $s_1$, is used to examine the degree of linear polarisation in the pulses, where:  $s_1=(E_x|^2-|E_y|^2)/(|E_x|^2+|E_y|^2)$ is the intensity difference between the $x$ and $y$ polarised fields normalised to the total intensity of the field. Values of $s_1 = \pm 1.0$ then indicate fully linear $x$ or $y$ polarisation respectively. This is plotted as a function of time in Fig.~\ref{fig:Stokes} where it is seen that the polarisation is highly modulated, flipping between the two polarisation states. The high degree of polarisation contrast is seen at the peak powers is $|s_1|>0.95$ demonstrating a high degree of polarisation modulation.

%Across the full radiation field $s1$ is consistently better than 0.9 with the higher power spikes giving the best degree of polarisation     

%shows the polarisation of the radiation field flipping between the different polarisation forms. 

%Pulses with polarisation alternating between left and right-hand polarisation is also modeled. For this case, the undulators in the afterburner are exchanges from plannar to helical. The amplifier section remains an x-polarised undulator. The power growth is inhibited by the energy modulation in the amplifier section and exponential in the afterburner as overlap between the radiation and areas of high beam quality is maintained. This allows a large ratio between the radiation produced in the afterburner and the amplifier. The extraction point of the electrons from the undulator is chosen such that it is 2 orders of magnitude smaller than the finial radiation field. Figure. \ref{fig:CircularPolPowerSpectrum} shows the power profiles for the left and right hand polarisation. 

Pulses with polarisation alternating between left and right-hand circular polarisation have also been modeled. For this case, the plane wave approximation was applied in simulations. It is expected that full 3D simulations will generate similar results as there was good agreement between the approximated linear polarisation case and the 3D results presented above. Full 3D simulations of alternating circular polarisation pulses will be subject of future study.  The amplifier section, which pre-bunches the electrons using SASE, remains an ($x$-polarised) planar undulator similar to that above. The afterburner now consists of orthogonal left and right circularly polarised helical undulators. 

%The power growth is inhibited by the energy modulation in the amplifier section and exponential in the afterburner due to the maintenance of overlap between the radiation and areas of high beam quality. This allows a large ratio between the radiation produced in the afterburner and the amplifier. The point at which the electron beam is extracted from the amplifying section is chosen at a position where the radiation is two orders of magnitude smaller than the final radiation field. 

\begin{figure}
    \centering
    \includegraphics[width=0.5\textwidth]{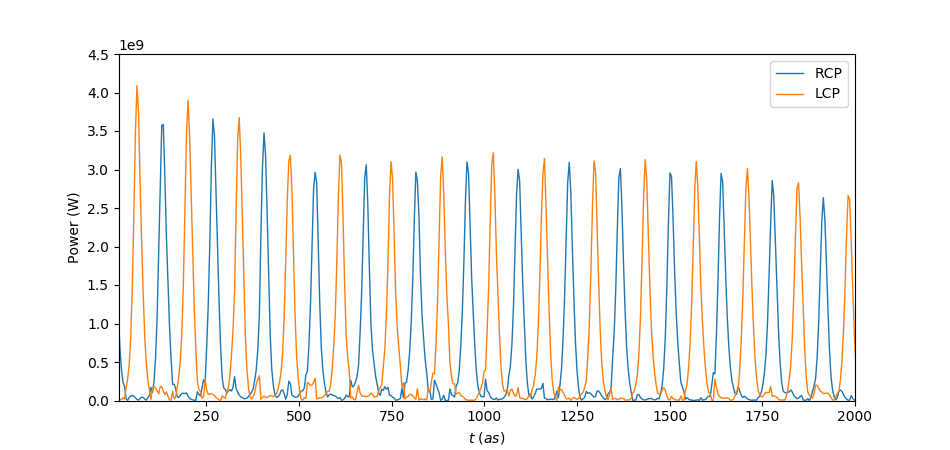}
    \caption{Power Spectrum for left and right hand polarisation vs relative time at the end an afterburner with alternating polarised helical undulator modules. }
    \label{fig:CircularPolPowerSpectrum}
\end{figure}

Fig.~\ref{fig:CircularPolPowerSpectrum} shows the power profiles for the left-hand circular, LCP, and right-hand circular, RCP, polarisation. The pulses now alternate between orthogonal circular polarisation with the same FWHM pulse duration $\tau_p$ and rate as the linearly polarised case above. At the pulse peaks, there is a high degree of circular polarisation, $|s_3|>0.9$, where $s_3=(|E_R|^2-|E_L|^2)/(|E_R|^2+|E_L|^2)$. This is very promising as most ultra-fast polarisation switching techniques cannot achieve full-handedness reversal.

%We have demonstrated attosecond polarisation modulation at soft X-rays. This has improved on the switching time scales currently ADD. Soft X-rays are no means the limit of that available with this scheme and the switching rate will increase as wavelengths shorten. Extrapolating from the simulation of the hard-Xray mode-locked afterburner presented in \cite{dunning2013few} the same parameters adapted to generate alternating polarised pulses would produce pulse separation times of $5$ as. Discussion of scaling the mode-locked afterburner to yet higher photon energies is provided in \cite{dunning2013few} would also apply to this alternating pulse set up. 

This Letter demonstrates a novel method to generate attosecond polarisation modulation in a short wavelength FEL without the need for any optical components. This represents a considerable improvement in wavelength and timescales over any other methods currently available, and could be expected to drive forward new experimental opportunities in fundamental science. The simulation parameters used considered soft X-ray pulses similar to the LCLS-II, however, this is by no means the limit of the wavelengths available with this set up. Extrapolating from the simulations of a hard X-ray mode-locked afterburner as presented in~\cite{dunning2013few}, the same parameters adapted to generate alternating polarised pulses here would generate pulse separation times of 5 as, approximately one fifth of the atomic unit of time. Discussion of scaling the mode-locked afterburner to yet higher photon energies provided in~\cite{dunning2013few} should also apply to the methods described here. Given the broad scaling of FEL wavelength operation, the method described will also be applicable to longer wavelengths, again opening up new experimental opportunities.

%Experimental implementation of mode-locking has not yet been trailed. It may be advantagous to consider a alternating pulse structure capablities when upgrading FEL to include mode-locking. 

%Adabtet to meet specific experimental requirements 

%Promising avenue to 

%Further modification of the set up is also possible. The examples in this letter show the case generating equal spacing between pulses. The temporal shift between the orthogonal polarised pulse trains is controlled by the slippage between them. It is therefore possible shorten to produce trains or two alternating pulses close together followed by a longer time interval under the condition that the slippage between undulator of the same type is $\lambda_m$.

As well as operating across a broad range of wavelengths, the method could be adapted to meet other specific experimental requirements. The temporal shift between pulse trains of orthogonal polarisations may be controlled to bring alternating pulses close together followed by a longer time interval. The time duration of the different pulse types may also be altered by the length of the different types of undulators to generate pulse trains with a pulse of one polarisation followed by a shorter pulse with the orthogonal polarisation. However, it is noted that this will also result in different pulse powers and bandwidths which would need further consideration. 

This method also provides a promising broader avenue to tailor FEL output and provide bespoke radiation for experiments. Further development of the method to be investigated will be to include alternating other pulse properties such as the wavelength, e.g. using the work of~\cite{2colour}, or orbital angular momentum of the pulses~\cite{OAM}.  While experimental implementation of mode-locking has not yet been trialed, it may be advantageous to consider alternating pulse structure capabilities when upgrading FELs to include mode-locking. 

We are grateful to funding from the Science and Technology Facilities Council (Agreement Number 4163192 Release\#3); ARCHIEWeSt HPC, EPSRC grant EP/K000586/1; EPSRC Grant EP/M011607/1; John von Neumann Institute for Computing (NIC) on JUROPA at J\"ulich Supercomputing Centre (JSC), project HHH20.

%\clearpage

%\bibliographystyle{ieeetr}

%\bibliography{alternatingpulses.bib}

\begin{thebibliography}{18}%
\makeatletter
\providecommand \@ifxundefined [1]{%
 \@ifx{#1\undefined}
}%
\providecommand \@ifnum [1]{%
 \ifnum #1\expandafter \@firstoftwo
 \else \expandafter \@secondoftwo
 \fi
}%
\providecommand \@ifx [1]{%
 \ifx #1\expandafter \@firstoftwo
 \else \expandafter \@secondoftwo
 \fi
}%
\providecommand \natexlab [1]{#1}%
\providecommand \enquote  [1]{``#1''}%
\providecommand \bibnamefont  [1]{#1}%
\providecommand \bibfnamefont [1]{#1}%
\providecommand \citenamefont [1]{#1}%
\providecommand \href@noop [0]{\@secondoftwo}%
\providecommand \href [0]{\begingroup \@sanitize@url \@href}%
\providecommand \@href[1]{\@@startlink{#1}\@@href}%
\providecommand \@@href[1]{\endgroup#1\@@endlink}%
\providecommand \@sanitize@url [0]{\catcode `\\12\catcode `\$12\catcode
  `\&12\catcode `\#12\catcode `\^12\catcode `\_12\catcode `\%12\relax}%
\providecommand \@@startlink[1]{}%
\providecommand \@@endlink[0]{}%
\providecommand \url  [0]{\begingroup\@sanitize@url \@url }%
\providecommand \@url [1]{\endgroup\@href {#1}{\urlprefix }}%
\providecommand \urlprefix  [0]{URL }%
\providecommand \Eprint [0]{\href }%
\providecommand \doibase [0]{http://dx.doi.org/}%
\providecommand \selectlanguage [0]{\@gobble}%
\providecommand \bibinfo  [0]{\@secondoftwo}%
\providecommand \bibfield  [0]{\@secondoftwo}%
\providecommand \translation [1]{[#1]}%
\providecommand \BibitemOpen [0]{}%
\providecommand \bibitemStop [0]{}%
\providecommand \bibitemNoStop [0]{.\EOS\space}%
\providecommand \EOS [0]{\spacefactor3000\relax}%
\providecommand \BibitemShut  [1]{\csname bibitem#1\endcsname}%
\let\auto@bib@innerbib\@empty
%</preamble>
\bibitem [{\citenamefont {S\'anchez-Barriga}\ \emph {et~al.}(2016)\citenamefont
  {S\'anchez-Barriga}, \citenamefont {Golias}, \citenamefont {Varykhalov},
  \citenamefont {Braun}, \citenamefont {Yashina}, \citenamefont {Schumann},
  \citenamefont {Min\'ar}, \citenamefont {Ebert}, \citenamefont {Kornilov},\
  and\ \citenamefont {Rader}}]{PhysRevB.93.155426}%
  \BibitemOpen
  \bibfield  {author} {\bibinfo {author} {\bibfnamefont {J.}~\bibnamefont
  {S\'anchez-Barriga}}, \bibinfo {author} {\bibfnamefont {E.}~\bibnamefont
  {Golias}}, \bibinfo {author} {\bibfnamefont {A.}~\bibnamefont {Varykhalov}},
  \bibinfo {author} {\bibfnamefont {J.}~\bibnamefont {Braun}}, \bibinfo
  {author} {\bibfnamefont {L.~V.}\ \bibnamefont {Yashina}}, \bibinfo {author}
  {\bibfnamefont {R.}~\bibnamefont {Schumann}}, \bibinfo {author}
  {\bibfnamefont {J.}~\bibnamefont {Min\'ar}}, \bibinfo {author} {\bibfnamefont
  {H.}~\bibnamefont {Ebert}}, \bibinfo {author} {\bibfnamefont
  {O.}~\bibnamefont {Kornilov}}, \ and\ \bibinfo {author} {\bibfnamefont
  {O.}~\bibnamefont {Rader}},\ }\href {\doibase 10.1103/PhysRevB.93.155426}
  {\bibfield  {journal} {\bibinfo  {journal} {Phys. Rev. B}\ }\textbf {\bibinfo
  {volume} {93}},\ \bibinfo {pages} {155426} (\bibinfo {year}
  {2016})}\BibitemShut {NoStop}%
\bibitem [{\citenamefont {Nicholls}\ \emph {et~al.}(2017)\citenamefont
  {Nicholls}, \citenamefont {Rodr{\'\i}guez-Fortu{\~n}o}, \citenamefont
  {Nasir}, \citenamefont {C{\'o}rdova-Castro}, \citenamefont {Olivier},
  \citenamefont {Wurtz},\ and\ \citenamefont {Zayats}}]{nicholls2017ultrafast}%
  \BibitemOpen
  \bibfield  {author} {\bibinfo {author} {\bibfnamefont {L.~H.}\ \bibnamefont
  {Nicholls}}, \bibinfo {author} {\bibfnamefont {F.~J.}\ \bibnamefont
  {Rodr{\'\i}guez-Fortu{\~n}o}}, \bibinfo {author} {\bibfnamefont {M.~E.}\
  \bibnamefont {Nasir}}, \bibinfo {author} {\bibfnamefont {R.~M.}\ \bibnamefont
  {C{\'o}rdova-Castro}}, \bibinfo {author} {\bibfnamefont {N.}~\bibnamefont
  {Olivier}}, \bibinfo {author} {\bibfnamefont {G.~A.}\ \bibnamefont {Wurtz}},
  \ and\ \bibinfo {author} {\bibfnamefont {A.~V.}\ \bibnamefont {Zayats}},\
  }\href {\doibase 10.1038/s41566-017-0002-6} {\bibfield  {journal} {\bibinfo
  {journal} {Nature Photonics}\ }\textbf {\bibinfo {volume} {11}},\ \bibinfo
  {pages} {628} (\bibinfo {year} {2017})}\BibitemShut {NoStop}%
\bibitem [{\citenamefont {Bull}\ \emph {et~al.}(2004)\citenamefont {Bull},
  \citenamefont {Jaeger}, \citenamefont {Kato}, \citenamefont {Fairburn},
  \citenamefont {Reid},\ and\ \citenamefont {Ghanipour}}]{10.1117/12.567640}%
  \BibitemOpen
  \bibfield  {author} {\bibinfo {author} {\bibfnamefont {J.~D.}\ \bibnamefont
  {Bull}}, \bibinfo {author} {\bibfnamefont {N.~A.}\ \bibnamefont {Jaeger}},
  \bibinfo {author} {\bibfnamefont {H.}~\bibnamefont {Kato}}, \bibinfo {author}
  {\bibfnamefont {M.}~\bibnamefont {Fairburn}}, \bibinfo {author}
  {\bibfnamefont {A.}~\bibnamefont {Reid}}, \ and\ \bibinfo {author}
  {\bibfnamefont {P.}~\bibnamefont {Ghanipour}},\ }in\ \href {\doibase
  10.1117/12.567640} {\emph {\bibinfo {booktitle} {Photonics North 2004:
  Optical Components and Devices}}},\ Vol.\ \bibinfo {volume} {5577},\ \bibinfo
  {editor} {edited by\ \bibinfo {editor} {\bibfnamefont {J.~C.}\ \bibnamefont
  {Armitage}}, \bibinfo {editor} {\bibfnamefont {S.}~\bibnamefont {Fafard}},
  \bibinfo {editor} {\bibfnamefont {R.~A.}\ \bibnamefont {Lessard}}, \ and\
  \bibinfo {editor} {\bibfnamefont {G.~A.}\ \bibnamefont {Lampropoulos}}},\
  \bibinfo {organization} {International Society for Optics and Photonics}\
  (\bibinfo  {publisher} {SPIE},\ \bibinfo {year} {2004})\ pp.\ \bibinfo
  {pages} {133 -- 143}\BibitemShut {NoStop}%
\bibitem [{\citenamefont {Yang}\ \emph {et~al.}(2017)\citenamefont {Yang},
  \citenamefont {Kelley}, \citenamefont {Sachet}, \citenamefont {Campione},
  \citenamefont {Luk}, \citenamefont {Maria}, \citenamefont {Sinclair},\ and\
  \citenamefont {Brener}}]{yang2017femtosecond}%
  \BibitemOpen
  \bibfield  {author} {\bibinfo {author} {\bibfnamefont {Y.}~\bibnamefont
  {Yang}}, \bibinfo {author} {\bibfnamefont {K.}~\bibnamefont {Kelley}},
  \bibinfo {author} {\bibfnamefont {E.}~\bibnamefont {Sachet}}, \bibinfo
  {author} {\bibfnamefont {S.}~\bibnamefont {Campione}}, \bibinfo {author}
  {\bibfnamefont {T.~S.}\ \bibnamefont {Luk}}, \bibinfo {author} {\bibfnamefont
  {J.-P.}\ \bibnamefont {Maria}}, \bibinfo {author} {\bibfnamefont {M.~B.}\
  \bibnamefont {Sinclair}}, \ and\ \bibinfo {author} {\bibfnamefont
  {I.}~\bibnamefont {Brener}},\ }\href {\doibase 10.1038/nphoton.2017.64}
  {\bibfield  {journal} {\bibinfo  {journal} {Nature Photonics}\ }\textbf
  {\bibinfo {volume} {11}},\ \bibinfo {pages} {390} (\bibinfo {year}
  {2017})}\BibitemShut {NoStop}%
\bibitem [{\citenamefont {Holldack}\ \emph {et~al.}(2020)\citenamefont
  {Holldack}, \citenamefont {Sch{\"u}ssler-Langeheine}, \citenamefont
  {Goslawski}, \citenamefont {Pontius}, \citenamefont {Kachel}, \citenamefont
  {Armborst}, \citenamefont {Ries}, \citenamefont {Sch{\"a}licke},
  \citenamefont {Scheer}, \citenamefont {Frentrup} \emph
  {et~al.}}]{holldack2020flipping}%
  \BibitemOpen
  \bibfield  {author} {\bibinfo {author} {\bibfnamefont {K.}~\bibnamefont
  {Holldack}}, \bibinfo {author} {\bibfnamefont {C.}~\bibnamefont
  {Sch{\"u}ssler-Langeheine}}, \bibinfo {author} {\bibfnamefont
  {P.}~\bibnamefont {Goslawski}}, \bibinfo {author} {\bibfnamefont
  {N.}~\bibnamefont {Pontius}}, \bibinfo {author} {\bibfnamefont
  {T.}~\bibnamefont {Kachel}}, \bibinfo {author} {\bibfnamefont
  {F.}~\bibnamefont {Armborst}}, \bibinfo {author} {\bibfnamefont
  {M.}~\bibnamefont {Ries}}, \bibinfo {author} {\bibfnamefont {A.}~\bibnamefont
  {Sch{\"a}licke}}, \bibinfo {author} {\bibfnamefont {M.}~\bibnamefont
  {Scheer}}, \bibinfo {author} {\bibfnamefont {W.}~\bibnamefont {Frentrup}},
  \emph {et~al.},\ }\href {\doibase 10.1038/s42005-020-0331-5} {\bibfield
  {journal} {\bibinfo  {journal} {Communications Physics}\ }\textbf {\bibinfo
  {volume} {3}},\ \bibinfo {pages} {1} (\bibinfo {year} {2020})}\BibitemShut
  {NoStop}%
\bibitem [{\citenamefont {McNeil}\ and\ \citenamefont
  {Thompson}(2010)}]{nphoton}%
  \BibitemOpen
  \bibfield  {author} {\bibinfo {author} {\bibfnamefont {B.~W.}\ \bibnamefont
  {McNeil}}\ and\ \bibinfo {author} {\bibfnamefont {N.~R.}\ \bibnamefont
  {Thompson}},\ }\href {\doibase 10.1038/nphoton.2010.239} {\bibfield
  {journal} {\bibinfo  {journal} {Nature Photonics}\ }\textbf {\bibinfo
  {volume} {4}},\ \bibinfo {pages} {814} (\bibinfo {year} {2010})}\BibitemShut
  {NoStop}%
\bibitem [{\citenamefont {Bonifacio}\ \emph {et~al.}(1994)\citenamefont
  {Bonifacio}, \citenamefont {De~Salvo}, \citenamefont {Pierini}, \citenamefont
  {Piovella},\ and\ \citenamefont {Pellegrini}}]{SASE}%
  \BibitemOpen
  \bibfield  {author} {\bibinfo {author} {\bibfnamefont {R.}~\bibnamefont
  {Bonifacio}}, \bibinfo {author} {\bibfnamefont {L.}~\bibnamefont {De~Salvo}},
  \bibinfo {author} {\bibfnamefont {P.}~\bibnamefont {Pierini}}, \bibinfo
  {author} {\bibfnamefont {N.}~\bibnamefont {Piovella}}, \ and\ \bibinfo
  {author} {\bibfnamefont {C.}~\bibnamefont {Pellegrini}},\ }\href {\doibase
  10.1103/PhysRevLett.73.70} {\bibfield  {journal} {\bibinfo  {journal} {Phys.
  Rev. Lett.}\ }\textbf {\bibinfo {volume} {73}},\ \bibinfo {pages} {70}
  (\bibinfo {year} {1994})}\BibitemShut {NoStop}%
\bibitem [{\citenamefont {Lutman}\ \emph {et~al.}(2016)\citenamefont {Lutman},
  \citenamefont {MacArthur}, \citenamefont {Ilchen}, \citenamefont {Lindahl},
  \citenamefont {Buck}, \citenamefont {Coffee}, \citenamefont {Dakovski},
  \citenamefont {Dammann}, \citenamefont {Ding}, \citenamefont {D{\"u}rr} \emph
  {et~al.}}]{lutman2016polarization}%
  \BibitemOpen
  \bibfield  {author} {\bibinfo {author} {\bibfnamefont {A.~A.}\ \bibnamefont
  {Lutman}}, \bibinfo {author} {\bibfnamefont {J.~P.}\ \bibnamefont
  {MacArthur}}, \bibinfo {author} {\bibfnamefont {M.}~\bibnamefont {Ilchen}},
  \bibinfo {author} {\bibfnamefont {A.~O.}\ \bibnamefont {Lindahl}}, \bibinfo
  {author} {\bibfnamefont {J.}~\bibnamefont {Buck}}, \bibinfo {author}
  {\bibfnamefont {R.~N.}\ \bibnamefont {Coffee}}, \bibinfo {author}
  {\bibfnamefont {G.~L.}\ \bibnamefont {Dakovski}}, \bibinfo {author}
  {\bibfnamefont {L.}~\bibnamefont {Dammann}}, \bibinfo {author} {\bibfnamefont
  {Y.}~\bibnamefont {Ding}}, \bibinfo {author} {\bibfnamefont {H.~A.}\
  \bibnamefont {D{\"u}rr}},  \emph {et~al.},\ }\href {\doibase
  10.1038/nphoton.2016.79} {\bibfield  {journal} {\bibinfo  {journal} {Nature
  photonics}\ }\textbf {\bibinfo {volume} {10}},\ \bibinfo {pages} {468}
  (\bibinfo {year} {2016})}\BibitemShut {NoStop}%
\bibitem [{\citenamefont {Thompson}\ and\ \citenamefont
  {McNeil}(2008)}]{PhysRevLett.100.203901}%
  \BibitemOpen
  \bibfield  {author} {\bibinfo {author} {\bibfnamefont {N.~R.}\ \bibnamefont
  {Thompson}}\ and\ \bibinfo {author} {\bibfnamefont {B.~W.~J.}\ \bibnamefont
  {McNeil}},\ }\href {\doibase 10.1103/PhysRevLett.100.203901} {\bibfield
  {journal} {\bibinfo  {journal} {Phys. Rev. Lett.}\ }\textbf {\bibinfo
  {volume} {100}},\ \bibinfo {pages} {203901} (\bibinfo {year}
  {2008})}\BibitemShut {NoStop}%
\bibitem [{\citenamefont {Dunning}\ \emph {et~al.}(2013)\citenamefont
  {Dunning}, \citenamefont {McNeil},\ and\ \citenamefont
  {Thompson}}]{dunning2013few}%
  \BibitemOpen
  \bibfield  {author} {\bibinfo {author} {\bibfnamefont {D.~J.}\ \bibnamefont
  {Dunning}}, \bibinfo {author} {\bibfnamefont {B.~W.~J.}\ \bibnamefont
  {McNeil}}, \ and\ \bibinfo {author} {\bibfnamefont {N.~R.}\ \bibnamefont
  {Thompson}},\ }\href {\doibase 10.1103/PhysRevLett.110.104801} {\bibfield
  {journal} {\bibinfo  {journal} {Phys. Rev. Lett.}\ }\textbf {\bibinfo
  {volume} {110}},\ \bibinfo {pages} {104801} (\bibinfo {year}
  {2013})}\BibitemShut {NoStop}%
\bibitem [{\citenamefont {Hemsing}(2020)}]{hemsing2020coherent}%
  \BibitemOpen
  \bibfield  {author} {\bibinfo {author} {\bibfnamefont {E.}~\bibnamefont
  {Hemsing}},\ }\href {\doibase 10.1103/PhysRevAccelBeams.23.020703} {\bibfield
   {journal} {\bibinfo  {journal} {Phys. Rev. Accel. Beams}\ }\textbf {\bibinfo
  {volume} {23}},\ \bibinfo {pages} {020703} (\bibinfo {year}
  {2020})}\BibitemShut {NoStop}%
\bibitem [{\citenamefont {Siegman}(1986)}]{siegman1986lasers}%
  \BibitemOpen
  \bibfield  {author} {\bibinfo {author} {\bibfnamefont {A.~E.}\ \bibnamefont
  {Siegman}},\ }\href@noop {} {\bibfield  {journal} {\bibinfo  {journal} {Mill
  Valley, CA}\ }\textbf {\bibinfo {volume} {37}},\ \bibinfo {pages} {169}
  (\bibinfo {year} {1986})}\BibitemShut {NoStop}%
\bibitem [{\citenamefont {Campbell}\ and\ \citenamefont
  {McNeil}(2012)}]{campbell2012puffin}%
  \BibitemOpen
  \bibfield  {author} {\bibinfo {author} {\bibfnamefont {L.}~\bibnamefont
  {Campbell}}\ and\ \bibinfo {author} {\bibfnamefont {B.}~\bibnamefont
  {McNeil}},\ }\href {\doibase 10.1063/1.4752743} {\bibfield  {journal}
  {\bibinfo  {journal} {Physics of Plasmas}\ }\textbf {\bibinfo {volume}
  {19}},\ \bibinfo {pages} {093119} (\bibinfo {year} {2012})}\BibitemShut
  {NoStop}%
\bibitem [{\citenamefont {Schoenlein}(2015)}]{LCLSIIScience}%
  \BibitemOpen
  \bibfield  {author} {\bibinfo {author} {\bibfnamefont {R.~W.}\ \bibnamefont
  {Schoenlein}},\ }\href@noop {} {\emph {\bibinfo {title} {{New Science
  Opportunities Enabled by LCLS-II X-ray Lasers}}}},\ \bibinfo {type} {Tech.
  Rep.}\ \bibinfo {number} {SLAC-R-1053}\ (\bibinfo  {institution} {{SLAC
  National Accelerator Laboratory}},\ \bibinfo {year} {2015})\BibitemShut
  {NoStop}%
%%CITATION = SLAC-PUB-14639;%%
\bibitem [{\citenamefont {Clarke}\ \emph {et~al.}(2012)\citenamefont {Clarke},
  \citenamefont {Jones},\ and\ \citenamefont {Thompson}}]{Clarke:2012zzb}%
  \BibitemOpen
  \bibfield  {author} {\bibinfo {author} {\bibfnamefont {J.}~\bibnamefont
  {Clarke}}, \bibinfo {author} {\bibfnamefont {J.}~\bibnamefont {Jones}}, \
  and\ \bibinfo {author} {\bibfnamefont {N.}~\bibnamefont {Thompson}},\
  }\href@noop {} {\bibfield  {journal} {\bibinfo  {journal} {Conf. Proc. C}\
  }\textbf {\bibinfo {volume} {1205201}},\ \bibinfo {pages} {1759} (\bibinfo
  {year} {2012})}\BibitemShut {NoStop}%
\bibitem [{\citenamefont {Thompson}(2019)}]{thompson:fel2019-thp033}%
  \BibitemOpen
  \bibfield  {author} {\bibinfo {author} {\bibfnamefont {N.}~\bibnamefont
  {Thompson}},\ }in\ \href {\doibase doi:10.18429/JACoW-FEL2019-THP033} {\emph
  {\bibinfo {booktitle} {Proc. FEL'19}}},\ \bibinfo {series and number}
  {\bibinfo {series} {Free Electron Laser Conference}\ No.~\bibinfo {number}
  {39}}\ (\bibinfo  {publisher} {JACoW Publishing, Geneva, Switzerland},\
  \bibinfo {year} {2019})\ pp.\ \bibinfo {pages} {658--660}\BibitemShut
  {NoStop}%
\bibitem [{\citenamefont {Campbell}\ \emph {et~al.}(2014)\citenamefont
  {Campbell}, \citenamefont {McNeil},\ and\ \citenamefont {Reiche}}]{2colour}%
  \BibitemOpen
  \bibfield  {author} {\bibinfo {author} {\bibfnamefont {L.~T.}\ \bibnamefont
  {Campbell}}, \bibinfo {author} {\bibfnamefont {B.~W.~J.}\ \bibnamefont
  {McNeil}}, \ and\ \bibinfo {author} {\bibfnamefont {S.}~\bibnamefont
  {Reiche}},\ }\href {\doibase 10.1088/1367-2630/16/10/103019} {\bibfield
  {journal} {\bibinfo  {journal} {New Journal of Physics}\ }\textbf {\bibinfo
  {volume} {16}},\ \bibinfo {pages} {103019} (\bibinfo {year}
  {2014})}\BibitemShut {NoStop}%
\bibitem [{\citenamefont {Hemsing}\ \emph {et~al.}(2011)\citenamefont
  {Hemsing}, \citenamefont {Marinelli},\ and\ \citenamefont
  {Rosenzweig}}]{OAM}%
  \BibitemOpen
  \bibfield  {author} {\bibinfo {author} {\bibfnamefont {E.}~\bibnamefont
  {Hemsing}}, \bibinfo {author} {\bibfnamefont {A.}~\bibnamefont {Marinelli}},
  \ and\ \bibinfo {author} {\bibfnamefont {J.~B.}\ \bibnamefont {Rosenzweig}},\
  }\href {\doibase 10.1103/PhysRevLett.106.164803} {\bibfield  {journal}
  {\bibinfo  {journal} {Phys. Rev. Lett.}\ }\textbf {\bibinfo {volume} {106}},\
  \bibinfo {pages} {164803} (\bibinfo {year} {2011})}\BibitemShut {NoStop}%
\end{thebibliography}

%merlin.mbs apsrev4-1.bst 2010-07-25 4.21a (PWD, AO, DPC) hacked
%Control: key (0)
%Control: author (8) initials jnrlst
%Control: editor formatted (1) identically to author
%Control: production of article title (-1) disabled
%Control: page (0) single
%Control: year (1) truncated
%Control: production of eprint (0) enabled
\providecommand{\noopsort}[1]{}\providecommand{\singleletter}[1]{#1}%

\end{document}